\documentclass[final,5p,times,twocolumn,authoryear]{elsarticle}

\usepackage{amssymb}
\usepackage{amsthm}
\usepackage{amsmath}
\usepackage{lipsum}
\usepackage{float}
\usepackage{orcidlink}

\journal{Physics Letters B}

\begin{document}

\begin{frontmatter}

\title{Wormholes in the $f(R,L,T)$ theory of gravity}
\author[first]{P.H.R.S. Moraes$^{*}$}
    \affiliation[first]{organization={Laboratório de Física Teórica e Computacional (LFTC), Universidade Cidade de São Paulo (UNICID)},%Department and Organization
            addressline={Rua Galvão Bueno 868}, 
            city={São Paulo},
            postcode={01506-000}, 
            state={São Paulo},
            country={Brazil}}
\author[second]{A.S. Agrawal\orcidlink{0000-0003-4976-8769}}\author[second]{B. Mishra\orcidlink{0000-0001-5527-3565}}         
    \affiliation[second]{organization={Department of Mathematics, Birla Institute of
Technology and Science-Pilani},%Department and Organization
            addressline={Hyderabad Campus}, 
            city={Hyderabad},
            postcode={500078}, 
            state={Telangana},
            country={India}}
\begin{abstract}
Morris and Thorne developed wormhole solutions in the late 1980s when they discovered a recipe that wormholes must follow for travelers to cross them safely. They describe exotic matter as satisfying $-p_{r} > \rho$, where $p_{r}$ is the radial pressure and $\rho$ is the energy density of the wormhole. This is a notable characteristic of the General Relativity Theory. The current article discusses traversable wormhole solutions in $f(R, L, T)=R+\alpha L+\beta T$, with $\alpha$ and $\beta$ are model parameters. The wormhole solutions presented here satisfy the metric constraints of traversability while remarkably avoiding the exotic matter condition, indicating that $f(R, L, T)$ gravity wormholes can be filled with ordinary matter. The derived solutions for the shape function of the wormhole meet the required metric conditions. They exhibit behavior that is comparable to that of wormholes reported in earlier references, which is also the case for our solutions for the energy density of such objects.
\end{abstract}

\end{frontmatter}

%\tableofcontents

%% \linenumbers

%% main text

\section{Introduction}\label{sec:int}

Our current model of cosmology, named $\Lambda$CDM model, is replete of flaws, inconsistencies or incompleteness. Ironically, what baptizes the model are exactly among its greatest shortcomings. According to $\Lambda$CDM model (also referred to as standard model), the cosmological constant $\Lambda$ is the responsible for the observed acceleration of the recent universe expansion \citep{riess/1998,perlmutter/1999}, but it carries ``the worst theoretical prediction in the history of physics'' \citep{hobson/2006}, namely, the {\it cosmological constant problem} \citep{weinberg/1989}, which is the huge discrepancy between the theoretical and observational values of $\Lambda$. On the other hand, cold dark matter (referred to as CDM) is an eccentric kind of matter that does not interact electromagnetically and therefore cannot be seen, but is necessary to fill galaxies and clusters of galaxies in order to fit observations \citep{salucci/2019,ashman/1992,limousin/2022,montes/2019}. Despite several and costly attempts, no particle associated to dark matter was ever detected \citep{smith/1990,boehm/2004,munoz/2004,mayet/2016}. Considering that according to the latest results by the Planck Satellite \citep{planck_collaboration/2020}, $\sim95\%$ of the universe composition is in the form of dark energy and dark matter, the above issues regarding $\Lambda$, which plays the role of ``dark energy'' in $\Lambda$CDM model, and the persistent non-detection of dark matter, are critical to the perenniality of the model.

Not only do the aforementioned problems persist, $\Lambda$CDM model also includes the {\it missing satellites problem} \citep{klypin/1999,moore/1999} and the {\it core/cusp problem} \citep{dubinski/1991,walker/2011}, which are, respectively, the over-abundance of the predicted number of halo substructures compared to the number of observed satellite galaxies and the difference between the model theoretical prediction that halos should follow a universal, centrally peaked (cuspy) density profile, and observations of dwarf and low surface brightness galaxies, that indicate shallower (cored) profiles. For a review on the fundamental problems of the $\Lambda$CDM model, which include the aforementioned ones, but are not restricted to them, we recommend \citep{bull/2016}.

A promising possibility to evade the above shortcomings is to work with {\it Extended Gravity Theories} (EGTs). The main motivation for considering EGTs is the possibility of describing, for instance, the dark sector of the universe, from extra gravitational terms that are not contained in Einstein's General Theory of Relativity (GR), which sustains $\Lambda$CDM model. Below we present and discuss some EGT examples.

The most popular EGT nowadays is the so-called $f(R)$ gravity theory, in which $R$ stands for the Ricci scalar. An important review on the subject is Reference \citep{de_felice/2010}. Effectively, the $f(R)$ theory replaces $R$ in the Einstein-Hilbert action
\begin{equation}\label{i1}
    S=\frac{1}{16\pi}\int R\sqrt{-g}d^4x,
\end{equation}
in which $g$ is the metric determinant, by a function of $R$, namely $f(R)$ (in units such that the speed of light $c$ and Newton's gravitational constant $G_N$ are taken both as $1$). Naturally, by applying the variational principle in the $f(R)$ action, the resulting field equations will present extra terms when compared to GR field equations, namely, 
\begin{equation}\label{i2}
    G_{\mu\nu}=8\pi T_{\mu\nu},
\end{equation}
with $G_{\mu\nu}$ being the Einstein tensor and $T_{\mu\nu}$ the energy-momentum tensor. The extra terms may, in principle, play the role of dark energy and dark matter in a different perspective; not as exotic fluids permeating the universe, but as modifying gravity terms \citep{amendola/2007,nojiri/2006,capozziello/2005,capozziello/2012,bohmer/2008}. Nevertheless, a number of $f(R)$ gravity flaws and shortcomings has already been reported, as one can check \citep{joras/2011} and the very recent reference \citep{casado-turrion/2023}, for instance, among others. Wormhole solutions in $f(R)$ gravity theory \citep{Mishra/2022, Agrawal/2022} and in unimodular gravity theory \citep{Agrawal/2023} obtained recently.

A common alternative to the $f(R)$ theory shortcomings is the $f(R,T)$ theory \citep{harko/2011}, for which $T$ stands for the trace of the energy-momentum tensor. While the $f(R)$ gravity is mathematically motivated by the possibility of higher order terms on $R$ in (\ref{i1}), the $f(R,T)$ gravity maintains this consideration, but also considers material terms to appear in (\ref{i1}). 

Some important applications of the $f(R,T)$ gravity are the following. Wormhole solutions in $f(R,T)$ gravity were obtained in \citep{sahoo/2018,moraes/2019}.  A cosmological model was constructed from the simplest non-minimal geometry-matter coupling in $f(R,T)$ gravity in \citep{moraes/2017}. The quantum version of $f(R,T)$ cosmology was constructed by \citep{xu/2016}. 

Another interesting form of coupling geometry and matter in a gravity theory is through the $f(R,L)$ gravity \citep{harko/2010}, with $L$ being the matter lagrangian density\footnote{Usually, this theory of gravity is referred to as $f(R,L_m,T)$ gravity. Here, in the present article, for the sake of simplicity we will refer to the matter lagrangian density simply as $L$, instead of $L_m$.}.

Some applications of the $f(R,L)$ gravity are the following: the hydrostatic equilibrium configurations of compact stars were first derived in the theory by \citep{carvalho/2020}. Later, quark stars with $2.6$M$_\odot$ were attained in \citep{carvalho/2022}. A cosmological model evading the Big-Bang singularity was obtained within $f(R,L)$ gravity in \citep{goncalves/2023}.

Further geometry-matter coupling theory approaches are the following: a Palatini formulation of EGT with non-minimal geometry-matter coupling was proposed by \citep{harko/2011b}. A thermodynamic interpretation of geometry-matter coupling gravity models was given by \citep{harko/2011c}. Galactic rotation curves within non-minimal geometry-matter coupling gravity were constructed by \citep{harko/2010b} and matter density perturbations were studied in \citep{nesseris/2009}.

In a recent paper by \citep{haghani/2021}, a generalized geometry-matter coupling theory of gravity was proposed, namely the $f(R,L,T)$ gravity. As the name would suggest, this theory proposes an $f(R,L,T)$ function to substitute $R$ in Eq.(\ref{i1}). % (Einstein-Hilbert action) 

Here we are going to construct, for the first time in the literature, traversable wormhole solutions in the generalized geometry-matter coupling gravity. Traversable wormholes are a class of GR solutions that allow to rapid interstellar travel \citep{morris/1988}. Such traversable wormholes must have no horizons and, according to GR, the material that generates the wormhole curvature must disobey the energy conditions. 

Wormholes have not been observed so far, but attempts to do so have been proposed. For instance, \citep{paul/2020} numerically constructed images of thin accretion disks in rotating wormhole backgrounds. The results show dramatic differences between such images and those obtained for black hole backgrounds. A method for detecting wormholes surrounded by optically thin dust was proposed by \citep{ohgami/2015}. The ring-down waveform of gravitational waves and gravitational lensing were proposed as a tool for distinguishing between wormholes and black holes by \citep{nandi/2017}. %Other proposals for observing wormholes can be seen in \citep{simonetti/2021,dai/2019}. 

There is a particular interest in studying traversable wormholes in modified theories of gravity, which is the fact that the extra degrees of freedom of such theories can allow for wormholes to be filled by non-exotic matter, i.e., matter satisfying the energy conditions. This has been attained in braneworld models \citep{sengupta/2022}, Born-Infeld gravity \citep{shaikh/2018}, Einstein-Cartan theory \citep{bronnikov/2015}, and $f(R,T)$ theory \citep{moraes/2018}.
\section{Static wormholes}\label{sec:sw}
The static wormhole metric is given by the Morris-Thorne solution \citep{morris/1988} as 
\begin{equation}\label{cw1}
    ds^2=-e^{2\Phi(r)}dt^2+\frac{dr^2}{1-\frac{b(r)}{r}}+r^2(d\theta^2+\sin^2\theta d\phi^2),
\end{equation}
with $\Phi$ being the redshift function and $b$ the shape function. 

In order for the wormhole spatial geometry to tend to the appropriate asymptotically flat limit, $\Phi$ must be such that 
\begin{equation}\label{cw2}
    \lim_{r\rightarrow{\infty}}\Phi<\infty.
\end{equation}

The traversable wormhole must obey
\begin{eqnarray}
    b(r_0)=r_0,\label{cw3}\\
    b'(r_0)\leq1,\label{cw4}
\end{eqnarray}
at the wormhole throat $r_0$, with a prime indicating radial derivative throughout the paper. One must also have
\begin{equation}\label{cw5}
    b'<\frac{b}{r}
\end{equation}
and away from $r_0$,
\begin{equation}\label{cw6}
    b<r
\end{equation}
must be satisfied.

Moreover, the energy-momentum tensor of the wormhole is
\begin{equation}\label{cw8}
    T^\mu_\nu=\texttt{diag}[-\rho,p_r,p_t,p_t],  
\end{equation}
in which $\rho$ is the wormhole matter-energy density, $p_r$ is its radial pressure and $p_t$ its tangential pressure. 
\section{The $f(R,L, T)$ gravity}\label{sec:frl}
The $f(R,L,T)$ gravity formalism  starts from the action \citep{haghani/2021}
\begin{equation}\label{frl1}
    S=\int d^4xf(R,L, T)\sqrt{-g}+\int d^4xL \sqrt{-g},
\end{equation}
in units such that $8\pi G_N=c=1$, which will be assumed.

The variation of (\ref{frl1}) with respect to metric tensor $g^{\mu \nu}$ yields
\begin{eqnarray}\label{frl2}
    &&(R_{\mu\nu}+g_{\mu\nu}\Box-\nabla_\mu\nabla_\nu)f_R-\frac{1}{2}fg_{\mu\nu}=8\pi T_{\mu\nu}+\nonumber\\
    &&\frac{1}{2}(f_L+2f_T)(T_{\mu\nu}-Lg_{\mu\nu})+f_T\tau_{\mu\nu}
\end{eqnarray}
as the field equations of the theory, for which $R_{\mu\nu}$ is the Ricci tensor, $f_R\equiv \frac{\partial f}{\partial R}$, $f_{L}\equiv \frac{\partial f}{\partial L}$, $f_{T}\equiv \frac{\partial f}{\partial T}$ and $\tau_{\mu\nu}\equiv2g^{\alpha\beta}\frac{\partial^2L}{\partial g^{\mu\nu}\partial g^{\alpha\beta}}$.

In the simple case of an additive structure of the gravitational Lagrangian of the form $f(R, L, T)=R+\alpha L+\beta T$, with $\alpha$ and $\beta$ being constants, and by assuming $L=-\rho$, the field equations \eqref{frl2} take the form
\begin{equation}\label{frl3}
G_{\mu\nu}=\left(8\pi+\frac{\alpha}{2}+\beta\right) T_{\mu \nu}+\frac{\beta}{2}(2\rho+T)g_{\mu\nu}.  
\end{equation}
\section{Wormholes in the $f(R,L,T)$ gravity}\label{sec:whfrlt}
Below, we present the non-null components of the field equations above for the wormhole metric (\ref{cw1}) and energy-momentum tensor (\ref{cw8}):
\begin{equation}\label{whfrlt1}
    -\frac{b}{r^3}=\left(\eta+\frac{\beta}{2}\right)p_{r}+\frac{\beta}{2}(\rho+2p_t),    
\end{equation}
\begin{equation}\label{whfrlt2}
\frac{1}{2r^2}\left(\frac{b}{r}-b'\right)=\left(\eta+\beta\right)p_t+\frac{\beta}{2}(\rho+p_r),    
\end{equation}
\begin{equation}\label{whfrlt3}
    \frac{b'}{r^2}=\left(\eta +\frac{\beta}{2}\right)\rho+\frac{\beta}{2}(p_r +2p_{t}),   
\end{equation}
where we have assumed $\Phi(r)=\texttt{constant}$ and $\eta\equiv8\pi +\frac{\alpha}{2}+\beta$.

By manipulating the above equations, we can write the matter content of the wormhole as
\begin{equation}\label{whfrlt4}
    \rho=\frac{b'}{\eta  r^2}=\frac{\rho^{\text{(eff)}}}{\eta},    
\end{equation}
\begin{equation}\label{whfrlt5}
    p_r=-\frac{b}{\eta  r^3}=\frac{p_{r}^{\text{(eff)}}}{\eta},    
\end{equation}
\begin{equation}\label{whfrlt6}
    p_t=\frac{b-r b'}{2 \eta  r^3}=\frac{p_{t}^{\text{(eff)}}}{\eta},  
\end{equation}
for which the superscript ``eff'' stands for effective.
\section{Energy Conditions}
The existence of wormhole solutions in GR depends on the violation of the null energy condition (NEC) \citep{Raychaudhuri/1955, morris/1988, Curiel/2017}.

The NEC states that \citep{Visser/1995}
\begin{equation}
    \text{NEC} \Leftrightarrow T_{\mu \nu}^{(\text{eff})}k^{\mu}k^{\nu}\geq 0
\end{equation}
for any null vector $k^\mu$. In terms of the principle pressure,
\begin{equation}
    \text{NEC} \Leftrightarrow ~\forall ~j, ~~~~\rho^{(\text{eff})}+p_{j}^{(\text{eff})}\geq 0.
\end{equation}

The weak energy condition (WEC) states that 
\begin{equation}
    \text{WEC} \Leftrightarrow T_{\mu \nu}^{(\text{eff})}V^{\mu}V^{\nu}\geq 0
\end{equation}
for any time-like vector $V^\mu$. In terms of the principle pressure,
\begin{equation}
    \text{WEC} \Leftrightarrow \rho^{(\text{eff})}\geq0, ~~~~ \text{and} ~~~~\forall j, ~~~~\rho^{(\text{eff})}+p_{j}^{(\text{eff})}\geq 0.
\end{equation}

The strong energy condition (SEC) states that 
\begin{equation}
    \text{SEC} \Leftrightarrow \left(T_{\mu \nu}^{(\text{eff})}-\frac{T}{2}g_{\mu \nu}\right)V^{\mu}V^{\nu}\geq 0
\end{equation}
for any time-like vector. In terms of principle pressures,
\begin{equation}
    \text{SEC} \Leftrightarrow \forall j, ~~~~\rho^{(\text{eff})}+p_{j}^{(\text{eff})}\geq 0, ~~~~ \text{and} ~~~~ \rho^{(\text{eff})}+\sum p_{j}^{(\text{eff})}\geq 0
\end{equation}
 
Accordingly, $T_{\mu \nu}^{(\text{eff})}k^{\mu}k^{\nu}<0$, which represents the violation of NEC requirement, is given by  
\begin{equation}
  \rho^{\text{eff}}+p_{r}^{\text{eff}}=\eta (\rho +p_{r})<0.  
\end{equation}

In the context of GR, the flaring-out condition and the NEC are incompatible. To satisfy the flaring-out condition,  $G_{\mu \nu} k^{\mu}k^{\nu} < 0$ must hold, in which $k^{\mu}$ is an arbitrary null vector. This implies, from Einstein’s field equations (\ref{i2}), that $T_{\mu \nu} k^{\mu}k^{\nu} < 0$. Choosing a stress-energy tensor $T_{\mu \nu}$ of the form given in Eq. (\ref{cw8}) leads to the violation of NEC. 

However, the situation is different in modified theories of gravity, where $T^{eff}_{\mu \nu}$ includes contributions not only from the matter sector but also from the extra gravitational degrees of freedom. Thus, although the effective stress-energy tensor must satisfy $T^{eff}_{\mu \nu} k^{\mu}k^{\nu} < 0$ to fulfill the flaring-out condition, it is still possible for the matter stress-energy tensor to satisfy the NEC, i.e., $T_{\mu \nu} k^{\mu}k^{\nu} > 0$, provided that the extra gravitational contributions compensate the positive matter contributions \citep{Luis Rosa and Kull/2022}.

Using the flaring out condition $\frac{b'r-b}{b^{2}}<0$ and imposing the WEC, given by $\rho>0$ and $\rho +p_{r}>0$, one can find that the coupling parameter $\eta$ is limited to $\eta <0$, and write
\begin{equation}
    p_{r}=\frac{-r b'-b+\eta  \rho  r^3}{\eta  r^3},
\end{equation}
\begin{equation}
    b(r)=r_{0} \left(\frac{r_{0}}{r}\right)^{\frac{1}{\omega}},
\end{equation}
where $p_{r} = \omega \rho$.

With the condition $b'<b/r$, the allowed region for the parameter $\omega$ is restricted to $1\geq\omega\geq0$, in which we are also taking the causality limit into account. Moreover, we can write
\begin{eqnarray}
    p_{r}=\omega \rho =-\frac{C r^{-\frac{1}{\omega }-3}}{\eta },~~
    %{\color{blue}\rho+p_{r}=-\frac{C (\omega +1) r^{-\frac{1}{\omega }-3}}{\eta  \omega }},\nonumber \\
    p_{t}=\frac{C (\omega +1) r^{-\frac{1}{\omega }-3}}{2 \eta  \omega }
\end{eqnarray}
with $C=r_{0}^{1+\frac{1}{\omega}}$, from which we see that an isotropic traversable wormhole would have to be filled by $\omega=-1/3$ matter, which is out of the acceptable range for our solution (and $\omega<0$ matter is out of the scope of the present letter as we are dealing with modified gravity).

In Figure 1 below, it can be seem that for the considered values of the model parameters, traversable wormhole conditions \eqref{cw5} and \eqref{cw6} are satisfied. 
\begin{figure}[H]
    \centering
    \includegraphics[scale=0.5]{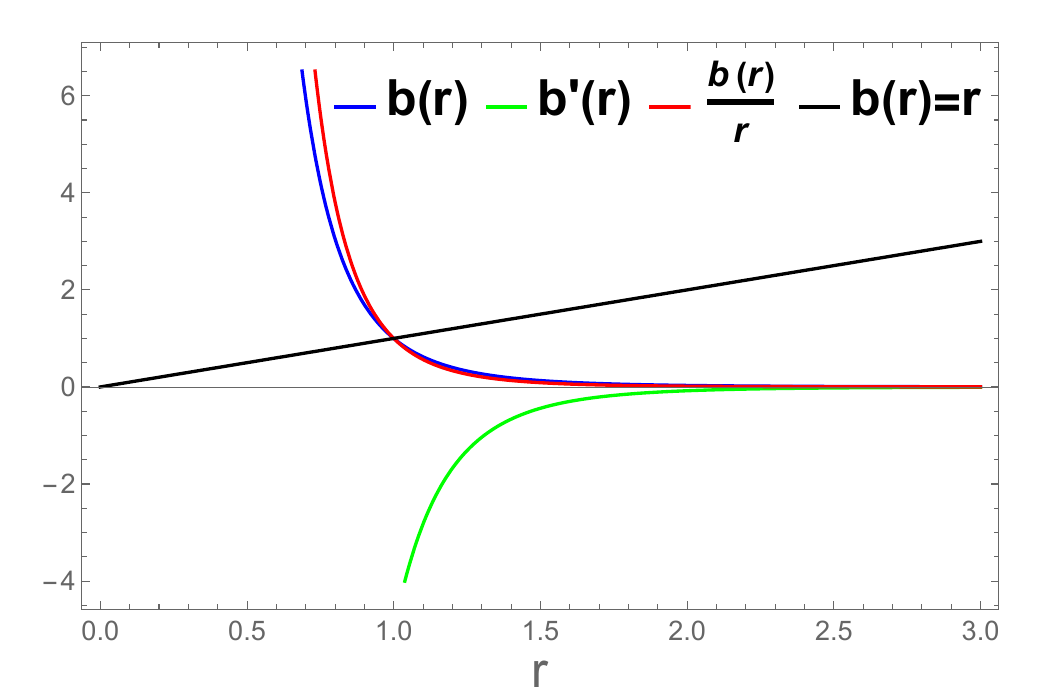}
    \caption{$b(r)$, $b'(r)$ and $b(r)/r$ versus
$r$ for $C=1$ and $\omega =0.2$.}
    \label{fig:enter-label2}
\end{figure}

Figures 2 and 3 below show that if the parameter $\omega\in (0,1)$, NEC is satisfied. 
\begin{figure}[H]
    \centering
    \includegraphics[scale=0.6]{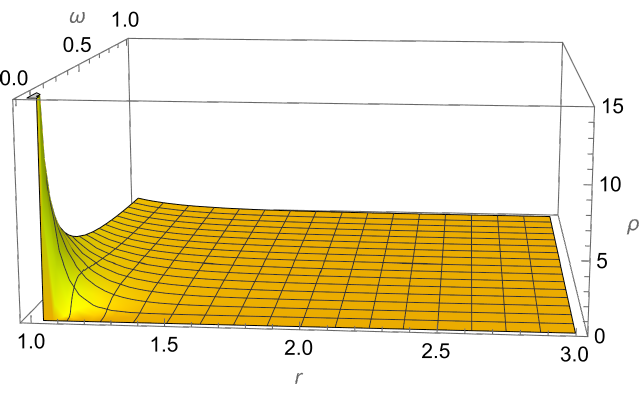}
    \caption{The energy density versus $r$ is represented for $\omega\in (0,1)$, with $C=1$ and $\eta =-1$.}
    \label{fig:enter-label3a}
\end{figure}
\begin{figure}[H]
    \centering
    \includegraphics[scale=0.7]{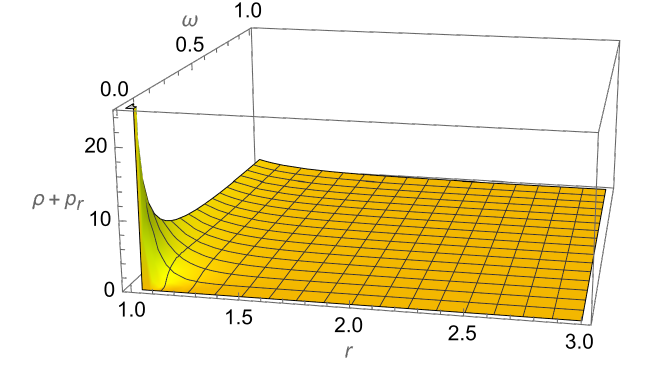}
    \includegraphics[scale=0.7]{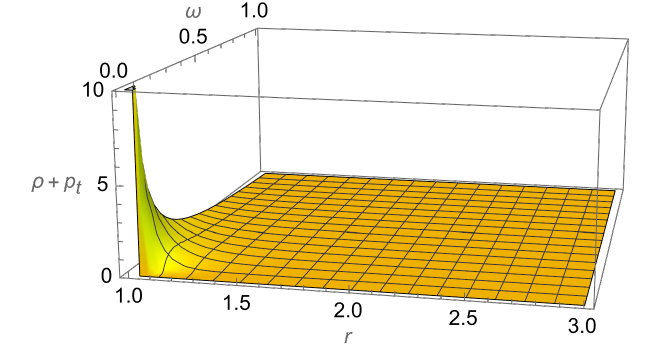}
    \caption{The null energy conditions versus
$r$ are represented for $\omega\in (0,1)$, with $C=1$ and $\eta =-1$.}
    \label{fig:enter-label5}
\end{figure}

Our solutions have shown that wormholes can be filled by dust, radiation and even stiff matter ($\omega=1$) \citep{Carr/2010, Bronnikov/2020} and still be traversable.
\section{Conclusions}

The traversable wormhole solution exists in GR and violates the NEC due to some form of exotic matter. But within the EGT context, the situation might be different. 

Here, the wormhole solution has been discussed in the $f(R, L, T)$ modified gravity. We have derived the gravitational field equations for this functional form. The shape function of the wormhole was then derived by specifying an equation of state for the matter threading it and applying the flaring out condition at the throat. We demonstrated that the stress-energy tensor of the matter threading the wormhole meets the NEC requirement in certain subspaces of the model parameter space. 
%For some values of $\omega$, WEC is also satisfied.

We have shown here that for the chosen functional form of the $f(R,L,T)$ function, it is possible to obtain traversable wormhole solutions satisfying energy conditions. 

Our present work can be extended in different fronts. Some other functional forms for the $f(R,L,T)$ function can be used to construct wormholes and check if those solutions are capable of describing traversable wormholes without exotic matter.

As the $f(R,L,T)$ gravity formalism was recently proposed, it still lacks several applications, such as those motivated in the Introduction. For instance, the same functional form used here can be extended to construct a cosmological model and check if for a range of values of $\alpha$ and $\beta$, it is possible to describe an accelerated expanding universe.

\section*{Acknowledgements} PHRSM would like to thank CNPq (Conselho Nacional de Desenvolvimento Científico e Tecnológico) for partial financial support. BM acknowledges the Council of Scientific and Industrial Research (CSIR) for the research grant to carry out the Research Project [No: 03/1493/23/EMR-II]. The Authors are thankful to the referee for bringing some important points to their attention. Those have enriched the physical content of the manuscript.

\end{document}